\documentclass[twocolumn]{emulateapj}

\usepackage{graphicx}
\usepackage{epsfig}

\shorttitle{Images of Spinning Neutron Stars}
\shortauthors{Baub\"ock et al.}

\begin{document}


\title{A Ray-Tracing Algorithm for Spinning Compact Object Spacetimes
  with Arbitrary Quadrupole Moments. II. Neutron Stars}

\author{Michi Baub\"ock, Dimitrios Psaltis, Feryal \"Ozel, and Tim Johannsen}
\affil{Astronomy Departments,
University of Arizona,
933 N.\ Cherry Avenue,
Tucson, AZ 85721, USA}
\email{email: mbaubock, dpsaltis, fozel, timj@email.arizona.edu}

\begin{abstract}
A moderately spinning neutron star acquires an oblate shape and a spacetime with a significant quadrupole moment. These two properties affect its apparent surface area for an observer at infinity, as well as the light curve arising from a hot spot on its surface. In this paper, we develop a ray-tracing algorithm to calculate the apparent surface areas of moderately spinning neutron stars making use of the Hartle--Thorne metric. This analytic metric allows us to calculate various observables of the neutron star in a way that depends only on its macroscopic properties and not on the details of its equation of state. We use this algorithm to calculate the changes in the apparent surface area, which could play a role in measurements of neutron-star radii and, therefore, in constraining their equation of state. We show that whether a spinning neutron star appears larger or smaller than its non-rotating counterpart depends primarily on its equatorial radius. For neutron stars with radii $\sim 10$~km, the corrections to the Schwarzschild spacetime cause the apparent surface area to increase with spin frequency. In contrast, for neutron stars with radii $\sim 15$~km, the oblateness of the star dominates the spacetime corrections and causes the apparent surface area to decrease with increasing spin frequency. In all cases, the change in the apparent geometric surface area for the range of observed spin frequencies is $\lesssim$ 5\% and hence only a small source of error in the measurement of neutron-star radii.
\end{abstract}


\section{INTRODUCTION}

Spectroscopic measurements of thermally emitting neutron-star surface areas have been widely used in attempts to infer the stellar radii and constrain the equation of state of ultradense matter (e.g., van Paradijs 1978, 1979; Rutledge et al. 2001; Majczyna \& Madej 2005; \"Ozel 2006). This approach has recently flourished with the use of high quality spectra obtained with current X-ray satellites during thermonuclear flashes (\"Ozel et al.\ 2009; G\"uver et al.\ 2010a, 2010b) and in the quiescent phases of accreting neutron stars (Heinke et al.\ 2006; Webb \& Barret 2007; Guillot et al.\ 2011). In some cases, the measured apparent surface areas possess formal uncertainties as small as a few percent (see, e.g., G\"uver et al.\ 2011) making it possible to distinguish between the predictions of different equations of state (\"Ozel et al.\ 2010; Steiner et al.\ 2010).

Converting a measurement of the apparent surface area of a neutron star to a radius requires correcting for a number of general-relativistic effects. As the photons propagate in the strong gravitational fields of neutron stars, their energies are redshifted while their trajectories experience strong lensing. For a non-spinning neutron star, both effects depend only on the $tt$-component of its exterior metric and can be analytically corrected for in any metric theory of gravity (Psaltis 2008). The situation, however, becomes increasingly complicated as the angular velocity of the neutron star increases toward the point of breakup.

The qualitative character of the general-relativistic effects depends on the ratio
\begin{equation}
\frac{f_{\rm s}}{f_0 }= 0.24 \left(\frac{f_{\rm s}}{600~{\rm Hz}}\right)\left(\frac{M}{1.8~M_\odot}\right)^{-1/2} \left(\frac{R}{10~\mathrm{km}}\right)^{3/2}
\label{omega}
\end{equation}
 of the spin frequency of the neutron star, $f_{\rm s}$, and the characteristic spin frequency $f_0=\sqrt{GM/R^3}/2\pi$ (Hartle \& Thorne 1968), where $G$ is the gravitational constant and $M$ and  $R$ are the mass and radius of the neutron star, respectively. To zeroth order in $f_{\rm s}/f_0$, the neutron star is spherically symmetric and its external spacetime is described by the Schwarzschild metric, which depends only on the mass of the star. To first order in $f_{\rm s}/f_0$, the neutron star remains spherically symmetric and its external spacetime is that of the Kerr metric. In this case, the amount of gravitational lensing depends also on the spin angular momentum of the star (because of the effects of frame dragging), which in turn depends on the density profile inside the star and hence on the equation of state. To second order in $f_{\rm s}/f_0$, the neutron star becomes oblate and its external spacetime is described by the Hartle--Thorne metric, which depends on the mass of the neutron star, on its spin angular momentum, and on its quadrupole mass moment. Finally, to even higher orders in $f_{\rm s}/f_0$, the external spacetime of the neutron star depends on multipole moments that are of increasing order. These spacetimes can be calculated numerically (see, e.g., Cook et al.\ 1994; Stergioulas \& Friedman 1995). In principle the external spacetime of a rapidly spinning neutron star can be accurately described by the analytic solution of Manko et al. (2000a, 2000b; see Berti \& Stergioulas 2004; Berti et al.\ 2005). However, the Manko et al.\ (2000a, 2000b) metric does not reduce to the Schwarzschild or Kerr solutions when the angular velocity of the star is reduced toward zero (Berti \& Stergioulas 2004), making it impractical for applications.

Various effects of increasing the angular velocity of a neutron star have been studied in the context of predicting the orbits of test particles (e.g., Shibata \& Sasaki 1998; Abramowicz et al.\ 2003; Berti \& Stergioulas 2004; Berti et al.\ 2005), the lightcurves that arise when the surface emission on a spinning neutron star is not uniform (Miller \& Lamb 1998; Braje et al.\ 2000; Muno et al.\ 2003; Poutanen \& Gierlinski 2003; Cadeau et al.\ 2005, 2007; Morsink et al.\ 2007), as well as the rotational broadening of atomic lines that originate on the stellar surfaces (\"Ozel \& Psaltis 2003; Bhattacharya et al.\ 2006; Chang et al.\ 2006). There are two conclusions that emerged from these studies and are relevant for our work: (1) that the Hartle-Thorne metric provides an approximation to the external spacetime of a rotating neutron star that is adequate for most astrophysical applications (Berti et al.\ 2005) and (2) that taking into account the oblateness of the stellar surface is at least as important as considering the effects of frame dragging around a spinning neutron star (Morsink et al.\ 2007).

In this article, we calculate numerically the apparent geometric surface area of a rotating neutron star using a variant of the Hartle--Thorne metric developed by Glampedakis \& Babak (2006). We use our new, fast ray-tracing algorithm that takes into account the oblateness of the neutron star as well as deviations from the Schwarzschild metric for its external spacetime that are formally correct up to the contributions of the quadrupole mass moment. With this analytic metric we can calculate the apparent surface area of a neutron star in a way that depends only on global properties of the star and not on the details of an assumed equation of state. Indeed, as we will show, the apparent surface area depends on six parameters: the equatorial and polar radii of the neutron star, its mass, spin angular momentum, and quadrupole mass moment, as well as the observer's inclination. Using the approximate relation between the polar and equatorial radii of spinning neutron stars obtained by Morsink et al.\ (2007) reduces the number of parameters to five.

We find that the apparent surface area of the neutron star changes only marginally over the range of spin frequencies of known sources. The corrections to the observed surface area only become significant at rotational frequencies greater than 1000~Hz. 

\section{The Spacetime of a Rapidly Spinning Neutron Star}

We describe the external spacetime of a rapidly spinning neutron star using the variant of the Hartle--Thorne metric developed by Glampedakis \& Babak (2006). This is based on the Kerr metric, but with the quadrupole moment allowed to deviate from its Kerr value. It therefore describes the spacetime of a rapidly spinning neutron star up to second order in $f_{\rm s}/f_0$. We opted to use this metric in order to contrast our results with those obtained when the external spacetime of a rapidly spinning neutron star is approximated by the Kerr metric. 

In Boyer--Lindquist coordinates, we write the metric as
\begin{equation}
g_{\mu\nu}=g_{\mu\nu}^{\rm K}+\eta a^2 h_{\mu\nu}\;.
\label{qKerr}
\end{equation}
Here $g_{\mu\nu}^{\rm K}$ is the Kerr metric that corresponds to the line element
\begin{eqnarray}
ds^2&=&-\left(1-\frac{2Mr}{\Sigma}\right)~dt^2
-\left(\frac{4Mar\sin^2\theta}{\Sigma}\right)~dtd\phi\nonumber\\
&&+\left(\frac{\Sigma}{\Delta}\right)~dr^2+\Sigma~d\theta^2\nonumber\\
&&+\left(r^2+a^2+\frac{2Ma^2r\sin^2\theta}{\Sigma}\right)\sin^2\theta~d\phi^2
\label{kerr}
\end{eqnarray}
with
\begin{equation}
\Delta\equiv r^2-2Mr+a^2,
\end{equation}
and
\begin{equation}
\Sigma\equiv r^2+a^2\cos^2~\theta\;.
\label{deltasigma}
\end{equation}
The quadrupole correction is given, in contravariant form, by
\begin{eqnarray}
h^{tt}&=&(1-2M/r)^{-1}\left[\left(1-3\cos^2\theta\right)
\mathcal{F}_1(r)\right],\nonumber\\
h^{rr}&=&(1-2M/r)\left[\left(1-3\cos^2\theta\right)\mathcal{F}_1(r)\right],
\nonumber\\
h^{\theta\theta}&=&-\frac{1}{r^2}\left[\left(1-3\cos^2\theta\right)
\mathcal{F}_2(r)\right],\nonumber\\
h^{\phi\phi}&=&-\frac{1}{r^2\sin^2\theta}\left[\left(1-3\cos^2\theta\right)
  \mathcal{F}_2(r)\right],\nonumber\\
h^{t\phi}&=&0\;,
\end{eqnarray}
with the functions $\mathcal{F}_{1,2}(r)$ shown explicitly in Appendix~A of Glampedakis \& Babak (2006). In these equations and hereafter, we set $G=c=1$. Note that, because it is of even order, the mass quadrupole affects only the diagonal components of the Kerr metric and not the $t\phi$-component that measures the amount of frame dragging. In this approach, the apparent surface area of a spinning neutron star depends on at least three macroscopic properties of the spacetime: the mass $M$, the specific angular momentum per unit mass, $a$, and the mass quadrupole moment $q$.

\begin{figure}[b]
\psfig{file=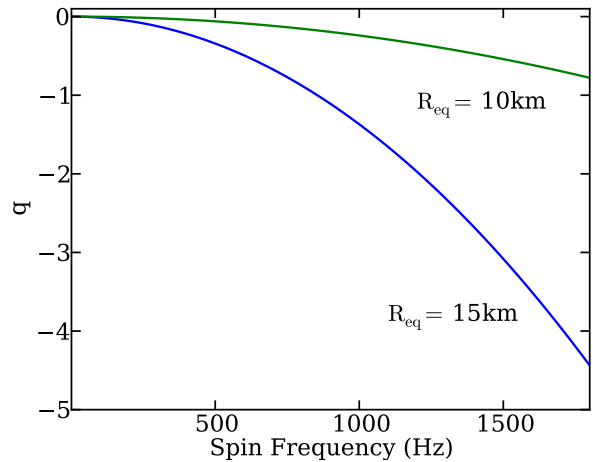,width=3.5in}
\caption{Mass quadrupole moment of the spacetime as a function of the spin frequency of the neutron star for two different radii. For each configuration, we show the quadrupole moment for a star of mass 1.8~$M_\odot$, with a radius of 10~km and 15~km, and the remaining parameters chosen to correspond to models with EOS FPS and EOS L, respectively (Cook et al.\ 1994). }
\label{fig:qplot}
\end{figure}

For a compact object with mass $M$ and specific angular momentum $a$, we write the mass quadrupole moment of the spacetime as
\begin{equation}
q=-a^2(1+\eta)\;,
\label{eq:quadrupole}
\end{equation}
so that, when $\eta=0$, this reduces to the quadrupole moment of the Kerr metric. (Note that, in the formalism of Glampedakis \& Babak 2006, $\epsilon=\eta a^2$.)  This mass moment of the spacetime for a rapidly spinning neutron star depends on its density profile and hence on the underlying equation of state. 

Laarakkers \& Poisson (1999) calculated neutron-star quadrupole moments for a wide range of equations of state, using a numerical algorithm that solves the Einstein field equations with no approximations. They found that Equation~(\ref{eq:quadrupole}) remains valid even for neutron stars that are spinning near their breakup points. They calculated $\eta\sim 1-6$, depending on the stellar mass and radius.  Berti \& Stergioulas (2004) further computed the quadrupole moments of stars in the Hartle--Thorne approximation and compared them to those of the numerical spacetimes. They concluded that the Hartle--Thorne approximation is adequate for all astrophysical applications. This approximation is expected to be valid for observed neutron stars because the fastest known X-ray burster is spinning at 620~Hz (4U 1608-52; Galloway et al. 2008), and the fastest known pulsar is spinning at 716~Hz (Hessels et al. 2006), which are expected to be significantly smaller than their breakup frequencies. Figure \ref{fig:qplot} shows the dependence of the quadrupole moment $q$ on the spin frequency for a 1.8~$M_\odot$ star with the appropriate value of the parameter $\eta$ taken from Laarakkers \& Poisson (1999) for two different equations of state. 

In order to calculate the apparent surface area of a neutron star using the above spacetime, we need to allow for the stellar surface to be non-spherical. This is required by the fact that the deviations of both the external spacetime from the Kerr metric as well as of the shape of the stellar surface from spherical symmetry are of second order in $f_{\rm s}/f_0$. Moreover, Morsink et al.\ (2007) have shown that, for purely geometric reasons, the changes in the predicted lightcurves of spinning neutron stars when the non-spherical shapes of their surfaces are taken into account are at least as important as the effects of frame dragging that are formally only of first order in $f_{\rm s}/f_0$.

\begin{figure}[t]
\psfig{file=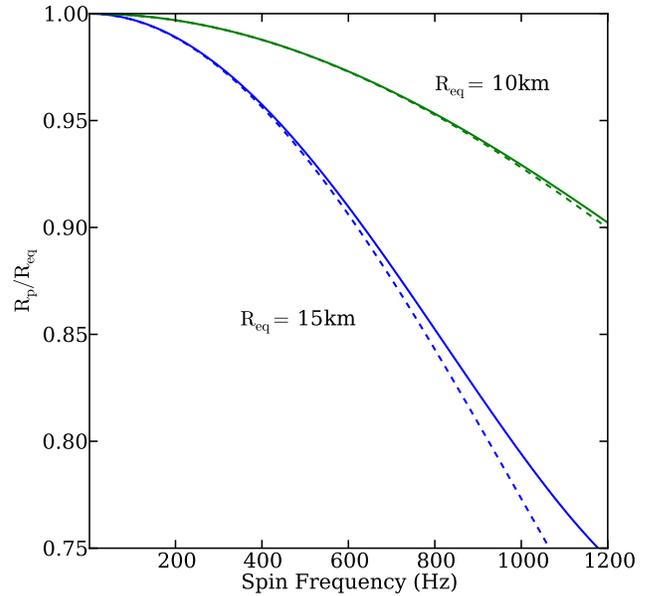,width=3.5in}
\caption{Ratio of the polar to the equatorial radius of a neutron   star as a function of its spin frequency, calculated using the   approximate relation of Morsink et al.\ (2007). The two pairs of curves correspond to two different stars with the same mass of 1.8~$M_\odot$ and different radii. In each case, the solid curve is the result of the complete fitting formula~(\ref{eq:shape}), whereas the dashed curve is the result when only the terms up to the quadrupole order are used.}
\label{fig:rpolar}
\end{figure}

Morsink et al.\ (2007) fit a large number of neutron-star shapes calculated for a wide range of equations of state. They show that an equation of the form
\begin{equation}
\frac{R(\theta)}{R_{\rm eq}}=1+\sum_{n=0}^N a_{2n}P_{2n}(\cos\theta),
\label{eq:shape}
\end{equation}
with $R_{\rm eq}$ the equatorial radius of the star and $P_{2n}(\cos\theta)$ the Legendre polynomial of order $2n$, accurately describes the shape of the neutron-star surface for even the fastest spinning stars if the series is terminated at $N=2$. For all equations of state the coefficients of the expansion depend only on two parameters,
\begin{equation}
\zeta\equiv \frac{GM}{R_{\rm eq}c^2}
\end{equation}
and
\begin{equation}
\epsilon\equiv \frac{(f_{\rm s}/2\pi)^2 R_{\rm eq}^2}{c^2 \zeta}\;.
\end{equation}
Excluding those equations of state that describe self-bound strange stars, Morsink et al.\ (2007) obtain for the coefficients in Equation~(\ref{eq:shape}) the following empirical relations:
\begin{eqnarray}
a_0&=&-0.18 \epsilon +0.23 \zeta \epsilon -0.05 \epsilon^2\nonumber\\
a_2&=&-0.39 \epsilon +0.29 \zeta \epsilon +0.13 \epsilon^2\nonumber\\
a_4&=& 0.04 \epsilon -0.15 \zeta \epsilon +0.07 \epsilon^2\;.
\end{eqnarray}
Figure~\ref{fig:rpolar} shows the ratio of the polar to the equatorial radius of a neutron star, as calculated with the above fitting formula.

In the Hartle--Thorne approximation, only terms up to second order are considered, i.e., the above series is terminated at $N=1$. In this case, we can rewrite the shape of the stellar surface in terms only of the equatorial and polar radii as
\begin{equation}
\frac{R(\theta)}{R_{\rm eq}}=\sin^2\theta +
\frac{R_{\rm p}}{R_{\rm eq}}\cos^2\theta\;.
\end{equation}
This introduces the two additional parameters, $R_{\rm p}$ and $R_{\rm eq}$, which will determine the apparent surface area of the neutron star. In Figure \ref{fig:rpolar}, we show the deformation of neutron stars for both the full fitting formula (solid lines) and the approximation up to quadrupole order (dashed lines).  Over the entire region of interest, the agreement between the approximation and the exact formula is within 5\%. Therefore disregarding higher order mass moment terms when calculating the shape of the neutron star is justified.

The last free parameter in our simulation is the inclination angle of the observer. Since there is no spherical symmetry, the apparent surface area of the neutron star will depend on the angle at which it is viewed. We define this angle, $\theta_0$, as the angle between the rotation axis of the star and the line of sight to an observer at infinity.

\section{A Numerical Algorithm for Ray Tracing}
\begin{figure}
\psfig{file=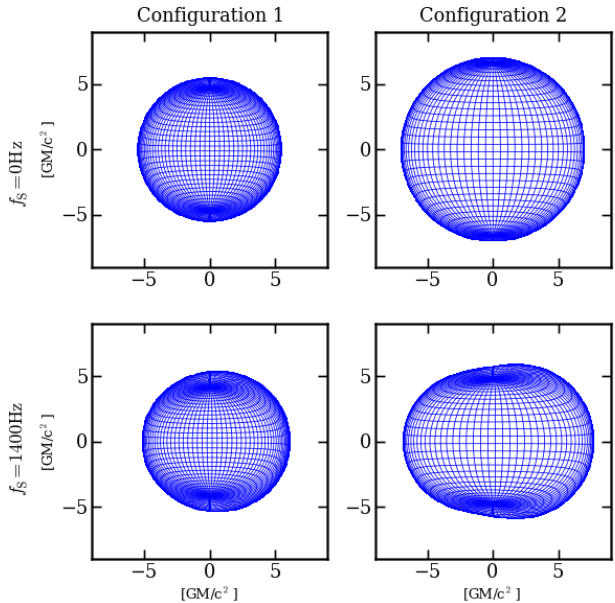, width=3.5in}
\caption{Apparent image of neutron stars at infinity calculated for different configurations. Two stars with $R_{\mathrm{eq}}$ = 15~km (left panels) and $R_{\mathrm{eq}}$ = 10~km (right panels) are shown, one non-spinning (upper panels) and one with a spin frequency of 1400~Hz (lower panels). In both cases, the observer is in the plane of the equator ($\theta_0 = \frac{\pi}{2}$), and the star rotates counterclockwise, the left side approaching the observer and the right side receding.}
\label{fig:Contours}
\end{figure}

\begin{figure*}
\psfig{file=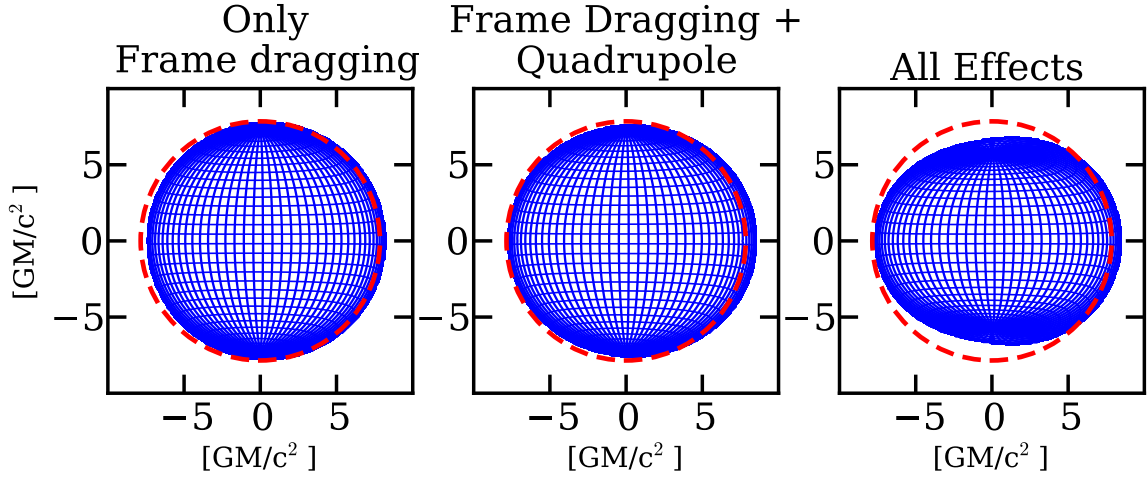, width=7in}
\caption{Effect of frame dragging, quadrupole moment, and oblateness on the image of a neutron star. The left panel shows the image of a star in a Kerr metric with $q=0$, where the only effect on the appearance of the star is the displacement of the image due to frame dragging. The center panel shows the star in a Kerr metric with  an extra quadrupole moment as described in Equation (\ref{eq:quadrupole}), which causes the image to stretch horizontally. The right panel includes both an extra quadrupole and an oblate stellar surface, causing the poles to squeeze together. The mass, radius, and moment of inertia for each star are chosen according to Configuration 2, as is the quadrupole moment for all but the Kerr case. In each case the star is rotating at 1000~Hz. The red dashed circle in each panel indicates the location and size of the image in the Schwarzschild metric.}
\label{fig:Distortion_Contours}
\end{figure*}

We follow the method outlined in Psaltis \& Johannsen (2011) to develop a ray-tracing algorithm that calculates the trajectories of photons in the neutron-star spacetime. The algorithm integrates the geodesic equations of the Hartle--Thorne metric using a Runge--Kutta solver. A number of rays originate perpendicular to an image plane at a large distance from the neutron star and are traced backward until they either intersect the surface of the star or exceed the distance to the star by 10\%, at which point we assume that they have bypassed the star entirely.

Because the surface of the neutron star deviates from spherical symmetry at high spin frequencies and photons may intersect the surface at oblique angles, special care must be taken when evaluating the coordinates at which the photon geodesics intersect the stellar surface. We use a simple procedure to determine the spherical coordinates of intersection: once a step in the Runge--Kutta algorithm crosses over the surface of the star, we perform a linear interpolation between the last two points on the geodesic. We then employ a bisection algorithm to find the point of intersection between that interpolation and the functional form of the stellar surface.

Following Cadeau et al.\ (2007), we calculate the beaming angle for a photon emitted at the surface of the neutron star as
\begin{equation}
\cos \alpha_e = \frac{\sqrt{g^{rr}}}{1+z} \frac{k^r - k^\theta R'(\theta)}{\sqrt{1 + [\frac{R'(\theta)}{R(\theta)}]^2}}
\label{eq:beamangle}
\end{equation}
where $k^r$ and $k^\theta$ are the components of the photon's momentum at the point of intersection with the stellar surface, $R(\theta)$ is the functional form of the surface of the neutron star (Equation (\ref{eq:shape})), and $z$ is the total redshift.

\section{Geometric Surface Areas of Neutron Stars}

One way of inferring the apparent surface area of the neutron star, in principle, is by direct angular measurement of the stellar image. If the distance of the star is known, it is straightforward to calculate its physical radius given its angular dimension. Another method to infer the surface area is by means of its thermal spectrum. By measuring the temperature of the star and its total flux, one can again derive its surface area. For slowly spinning neutron stars, these two measurements agree.

In the absence of rotation, the spacetime around the star is described by the Schwarzschild metric. An observer at infinity will observe the star to be enlarged due to gravitational self-lensing. The angular apparent surface area of the star is then given by
\begin{equation}
\frac{{A}_{\rm Sch}^G}{{A}} = (1-\frac{2GM}{R c^2})^{-1}\;,
\label{eq:rinfinity}
\end{equation}
where $A = \pi R^2$. The spectroscopic area, on the other hand, is defined as
\begin{equation}
A^S_{\rm Sch}= \frac{F_\infty}{\sigma T_{\mathrm{eff},\infty}^4},
\label{eq:aspec}
\end{equation}
where $F_\infty$ is the thermal flux and $T_{\mathrm{eff}, \infty}$ is its effective temperature. Gravitational redshift reduces both the observed flux and the effective temperature as 
\begin{eqnarray}
&&F_{\infty} =(1-\frac{2GM}{R c^2})F_\mathrm{NS}, \nonumber\\
&&T_{\infty} = \sqrt{1-\frac{2GM}{R c^2}} T_\mathrm{NS},
\label{eq:FandT}
\end{eqnarray}
where $F_\mathrm{NS}$ and $T_\mathrm{NS}$ are the flux and temperature measured at the surface of the neutron star. Combining Equations (\ref{eq:aspec}) and (\ref{eq:FandT}) we find that the apparent spectroscopic surface area at infinity is given by 
\begin{equation}
\frac{A_{\rm Sch}^S}{A} = 4 (1-\frac{2GM}{R c^2})^{-1}.
\label{eq:aspecainfty}
\end{equation}
Clearly, for a slowly spinning neutron star
\begin{equation}
\frac{A_{\rm Sch}^S}{A_{\rm Sch}^G} = 4.
\label{eq:ASAG}
\end{equation}
The factor of four arises from the difference between the total surface area ($4 \pi R^2$) measured spectroscopically and the projected surface area ($\pi R^2$) measured geometrically. For a moderately spinning neutron star, the lack of spherical symmetry in the metric, the oblateness of the star, and the second order terms in the Doppler shift will introduce different corrections to the geometric and spectroscopic areas as measured at infinity. As a result, these two measurements will not agree in general. 

In practice, only the spectroscopic area can be measured, as neutron stars are too small to allow for a direct measurement of their angular sizes. However, the change in apparent angular size of a neutron star plays a role even in the measurement of its spectroscopic surface area. In the following, we use our ray-tracing algorithm to investigate the effect of rapid rotation on the geometric surface areas of neutron stars. We will discuss the additional effects of position-dependent redshift and Doppler shifts in this metric in a forthcoming paper. 

\begin{figure*}
\psfig{file=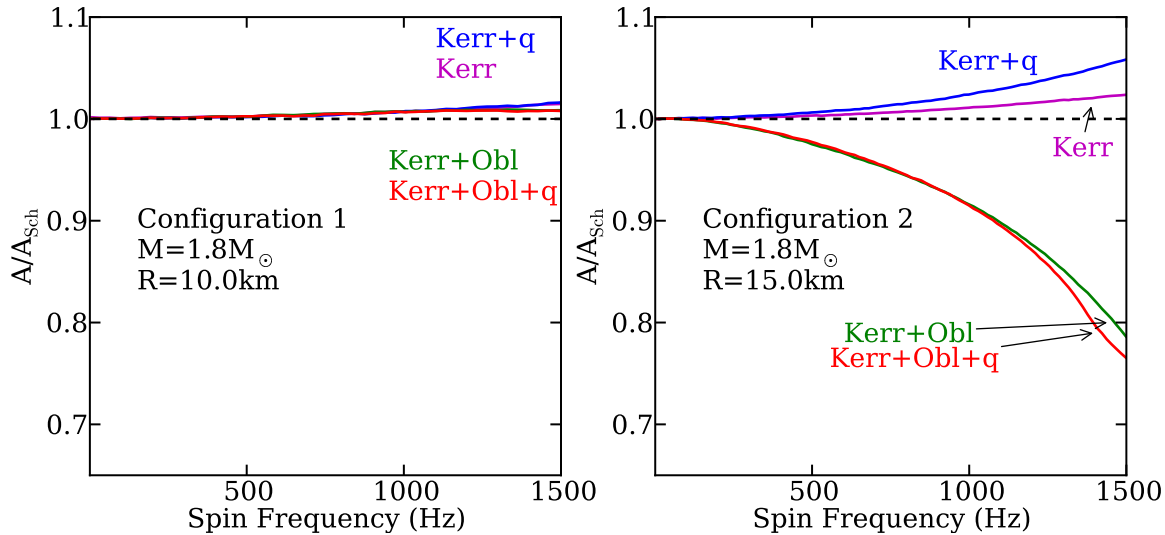, width=7in}
\caption{Ratio of the apparent surface area of a rotating neutron star to the area of a static neutron star as a function of spin frequency. Two stars are shown, corresponding to Configuration 1 (left panel) and Configuration 2 (right panel) as described in the text. In both plots, the four lines show different approximations of the metric outside the star: the Kerr metric, the Kerr metric plus a quadrupole deviation, the Kerr metric with an oblate stellar surface, and the Kerr metric with both a quadrupole deviation and oblateness. The dashed line is at $A/A_{\mathrm {Sch}} = 1$. In this plot, the inclination of the star is set to 90$^{\circ}$. Note that the plot is extended past the mass shed limit in some configurations in order to show the general trend of the relationship.}
\label{fig:mplot}
\end{figure*}

\begin{figure*}
\psfig{file=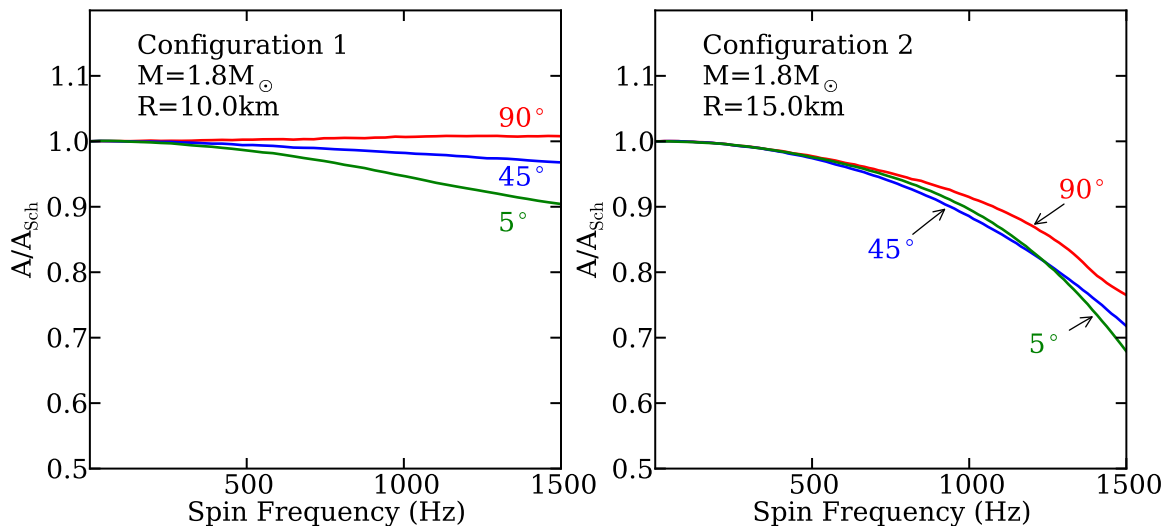, width=7in}
\caption{As in Figure~\ref{fig:mplot} but showing the apparent radius for three different inclinations, taking into account both the quadrupole moment and the oblateness of the star. The left panel corresponds to Configuration 1, while the right panel corresponds to Configuration 2. Again, the $x$-axis is extended past the mass-shed limit in some cases.}
\label{fig:incplot}
\end{figure*}

\begin{deluxetable}{lcccc}
\tablecolumns{5}
\tablecaption{Neutron-Star Parameters}
\tablehead{
 \colhead{Configuration} &
  \colhead{$M$}&
 \colhead{$R_{\mathrm{eq}}$} &
 \colhead{$I$} &
 \colhead{$\eta$}\cr
   &
 \colhead{$(M_\odot)$ } &
 \colhead{ (km)} &
 \colhead{$(10^{45} \mathrm{g} \mathrm{cm}^2)$}& 
}
\startdata
1 & 1.8 & 10 & 1.4& 1.2 \\
2 & 1.8 & 15  & 3.0 & 3.9
\enddata
\tablenotetext{}{Parameters of interest for the two neutron star configurations used in this paper: the mass $M$, the equatorial radius $R_{\mathrm{eq}}$, the moment of inertia $I$, and the deviation from the Kerr quadrupole $\eta$.}
\label{tb:nsparams}
\end{deluxetable}

We calculate the apparent geometric surface area of a neutron star by counting directly the number of rays originating at the image plane that intersect the surface of the star. The area on the image plane covered by the intersecting rays corresponds to the angular area of the image of the neutron star. Moreover, the coordinates on the neutron star at which each ray intersects the surface allow us to construct a contour plot of the apparent image of the star to an observer at infinity (see Figure~\ref{fig:Contours}).

For our models, we used two different configurations of neutron-star parameters, hereafter labeled ``Configuration 1" and ``Configuration 2". In each case, we chose the mass, equatorial radius, moment of inertia (from Cook et al.\ 1994), and quadrupole deviation $\eta$ (according to Laarakkers \& Poisson 1999) corresponding to a proposed equation of state in the non-spinning case. In order to demonstrate the range of the change in surface area, we chose a relatively soft equation of state (EOS FPS) for Configuration 1, and a relatively stiff equation of state (EOS L) for Configuration 2. We then held these parameters constant and varied only the spin frequency of the neutron star. Therefore, as the spin of each star increases, its parameters will increasingly deviate from the values predicted by the initial equations of state. While the resulting models do not reflect the predictions of any particular equation of state, they allow us to separate the various physical effects and estimate the magnitude of the change in surface area across the parameter space. The parameters of both configurations are summarized in Table~\ref{tb:nsparams}.  

Figure~\ref{fig:Contours} illustrates the appearance of the two neutron stars in both the Schwarzschild approximation and at a high spin frequency. Three effects influence the apparent image in the rapidly spinning case: frame dragging, the non-zero quadrupole moment, and the oblateness of the stellar surface. The change in the apparent image of the star due to each of these effects is illustrated in Figure~\ref{fig:Distortion_Contours}. Frame dragging causes the image of the neutron star to be displaced horizontally while preserving its circular appearance. The negative quadrupole moment causes the image to stretch horizontally about the axis of rotation. Finally, the oblateness of the stellar surface squeezes the poles together, further emphasizing the elongated appearance of the star. The combination of the displacement due to frame dragging and the stretching due to the quadrupole moment give rise to the asymmetric shape of the stellar image. A similar effect has been reported for images of accretion flows around black holes (Johannsen \& Psaltis 2010). 

Figure~\ref{fig:mplot} shows the contributions of each of these effects to the geometric surface area of the neutron star as a function of spin frequency. Note that the Hartle--Thorne approximation to the metric formally becomes inaccurate at the highest spin frequencies shown in this figure. The dominant effect on the image of a neutron star with a Kerr metric is due to frame dragging. However, as discussed above, frame dragging merely displaces the image of the star without altering its area. Therefore, the apparent area under the Kerr metric stays nearly constant (the small increase at high spin frequencies is due to the non-zero Kerr quadrupole). Including the extra quadrupole moment described by Equation~(\ref{eq:quadrupole}) increases the surface area more significantly at faster spins. The oblateness of the stellar surface tends to counteract these effects and decrease the apparent surface area. Whether the surface area increases or decreases with higher spin frequency depends on which of these corrections dominate.

Comparing the left and right panels of Figure~\ref{fig:mplot}, it is clear that the relative importance of the oblate shape of the star grows as its equatorial radius increases. Figure~\ref{fig:mplot} shows that the radius of a star described by Configuration 2 falls within the regime where the oblateness dominates the correction to the apparent surface area. In this case, the observed surface area of the star primarily decreases with increasing spin frequency, and the quadrupole moment only adds a small correction at the highest spin frequencies. At a radius of 10~km, however, as in Configuration 1, the effect of the quadrupole moment outweighs the oblateness for spin frequencies up to 1200~Hz, resulting in an overall increase of apparent surface area.

Since the presence of a spin axis breaks the symmetry of the spacetime, the image of the neutron star also depends on the inclination at which it is viewed. Figure~\ref{fig:incplot} shows this dependence for the two neutron stars with different radii. Again, the direction of the trend depends on which of the effects is dominant. For the neutron star with a larger radius (left panel), increasing the spin invariably leads to a smaller observed surface area and a weak dependence on inclination. For the neutron star with a smaller radius, however, the effects due to the oblateness and the quadrupole moment are comparable in magnitude (right panel). In this case, the direction of the effect depends more strongly on inclination: the star appears to grow in area with increasing spin frequency at large inclination, while it shrinks when viewed near the poles. 
\begin{figure}
\psfig{file=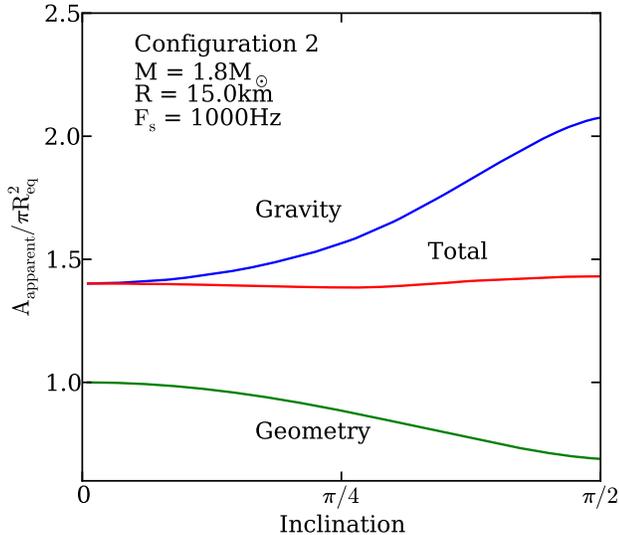,width=3.5in}
\caption{Relative contributions of geometry and gravitational light bending to the total surface area as a function of inclination. The green line, marked ``Geometry,'' shows the apparent surface area normalized to $\pi R_{\mathrm{eq}}^2$ in a flat spacetime (see the Appendix). The red line, labeled ``Total,'' shows the area calculated by ray tracing in the Hartle--Thorne metric. The ratio of these two areas shows the effect of gravitational light-bending as a function of inclination, plotted in the blue line labeled ``Gravity''.}
\label{fig:proj_plot}
\end{figure}

The change in surface area with inclination is due to the non-spherical shape of the neutron star and the $\theta$-dependence of the metric. We investigate the behavior of these two contributions further in Figure \ref{fig:proj_plot}, where we show the contributions of geometry and light-bending to this change in area for a fixed spin frequency for the Configuration 2 neutron star. 

The curve labeled ``Geometry'' shows the dependence of the apparent surface area of the neutron star on inclination in the absence of any relativistic effects (see the Appendix for the details of the calculation). As expected, changing the inclination from pole-on to edge-on decreases the surface area of the image. On the other hand, the curve labeled ``Total'' shows the change of the surface area with inclination when relativistic effects are taken into account. We also plot the ratio of the two dependences as the curve labeled ``Gravity.'' As the poles of the oblate spheroid lie deeper in the gravitational well than the equator, the effect of light-bending is stronger when the star is seen at a higher inclination. These two effects cancel to within 10\%, leaving the apparent surface area of the star practically constant over changing inclination.

\section{Conclusions}

A spinning neutron star introduces several effects in surface area measurements beyond those of the stationary Schwarzschild metric. Rapid rotation causes the neutron star to acquire a quadrupole moment and an oblate surface. Both of these effects distort the image of the star as seen by distant observers, introducing a correction to the calculation of neutron star radii from a measurement of their surface areas.

The relative contributions of the quadrupole moment compared to the oblateness depend strongly on the radius of the star. At a radius of 15~km, the quadrupole contribution is negligible when compared to the change in apparent area caused by the deformation of the stellar surface at all spin frequencies. At a radius of  10~km, which are consistent with current observations (\"Ozel et al.\ 2010; Steiner et al.\ 2010), the self-lensing dominates over the oblateness for spin frequencies less than $\sim1000$~Hz.

While the apparent surface area does depend on the oblateness and quadrupole moment caused by the neutron-star spin, these corrections are small for the spin frequencies that have been observed to date. Only at frequencies above $\sim800$~Hz does the apparent surface area change significantly. For the observed range of spin frequencies, the change in apparent geometric surface area is less than 10\%, for both configurations considered here and at all inclinations.

Studies of the spectroscopic determination of surface areas for thermally emitting neutron stars have so far relied on the Schwarzschild approximation, which neglects the correction due to the changes in spin and oblateness of the neutron star due to the stellar rotation (e.g., Rutledge et al.\ 2001; \"Ozel 2006; Heinke et al.\ 2006; Webb \& Barret 2007; \"Ozel et al.\ 2009; G\"uver et al.\ 2010a, 2010b). Observational uncertainties inherent in these measurements are currently at the $\sim 10\%$ level and hence larger than the effects presented here. We will discuss the additional corrections in the spectroscopic measurements of spinning neutron stars in a forthcoming paper. As the uncertainties in the data shrink with future observations, it will be necessary to take into account the corrections due to neutron star spins. 
 
 \acknowledgements
 
 We gratefully acknowledge support from Chandra Theory grant TMO-11003X, NSF CAREER award NSF 0746549, NASA ADAP grant NNX10AE89G, and NSF grant AST-1108753 for this work.

\appendix
 \renewcommand{\theequation}{A\arabic{equation}}
\gdef\thesection{Appendix \Alph{section}}
In this appendix we calculate the Newtonian surface area of the neutron star as a function of the inclination angle. Qualitatively, we are calculating the area of the shadow cast by the neutron star onto the image plane in the presence of light rays normal to the image plane. The outline of this figure is defined by the points on the neutron star at which a normal to the image plane is tangent to its surface.

To find this locus of points, we first write a parametric expression for a vector from the origin to a point on the surface of the neutron star with coordinates ($\theta$,$\phi$) in the usual form:
\begin{equation}
\mathbf{R}(\theta, \phi) = \{R(\theta)\sin(\theta)\cos(\phi), R(\theta)\sin(\theta)\sin(\phi), R(\theta)\cos(\theta)\},
\label{eq:rvec}
\end{equation}
with $R(\theta$) defined as in Equation (\ref{eq:shape}). We then define a normal vector to the surface at ($\theta$,$\phi$) as
\begin{equation}
\mathbf{N}(\theta, \phi) = \frac{\partial \mathbf{R}}{\partial \theta} \times \frac{\partial \mathbf{R}}{\partial \phi},
\label{eq:nvec}
\end{equation}
and a unit vector normal to the image plane at an angle $\theta_0$ to the rotational axis of the neutron star as $\mathbf{V}(\theta_0) = \{0, \sin(\theta_0), \cos(\theta_0)\}$. As $\theta_0$ varies from 0 to $\frac{\pi}{2}$, the inclination of the image plane ranges from pole-on to edge-on.

The boundary of the projection onto the image plane is defined by those points on the stellar surface which satisfy the equation
\begin{equation}
\mathbf{N}(\theta,\phi) \cdot \mathbf{V}(\theta_0) = 0.
\label{eq:taneq}
\end{equation}
This equation is linear in $\sin(\phi)$ and has the solution
\begin{equation}
\sin(\phi) = -\frac{\cos(\theta)\cot(\theta_0) R(\theta) + \cot(\theta_0) \sin(\theta) R'(\theta)}{R(\theta) \sin(\theta) - \cos(\theta)R'(\theta)}.
\label{eq:sinphi}
\end{equation}
This equation is no longer valid in the limiting cases $\theta = \frac{\pi}{2}$ and $\theta = 0$. In the former case, the outline of the neutron star is described by Equation (\ref{eq:shape}). In the latter case, when the observer lies along the rotational axis of the star, the outline will of course appear as a circle with radius $R(\frac{\pi}{2})$. It should also be noted that not all values of colatitude $\theta$ will correspond to a physical solution for the azimuth $\phi$---the projected circumference of the image will lie within a range of $\theta$ around the equator. We denote by $\theta_+$ and $\theta_-$ the limits of colatitude that contribute to the image at infinity.

In order to project the resulting outline of the neutron star onto the image plane, we simply rotate the coordinate system by an angle $-\theta_0$ so the $x$--$y$ plane is parallel to the image plane. We then set the $z$-component of all the points on the outline to zero to find a two-dimensional parametric form $\{C_x(\theta;\theta_0),C_y(\theta;\theta_0)\}$ of the outline of the projected image. Finally, we numerically integrate this curve to find the projected surface area:
\begin{equation}
A = 2 \int_{\theta_-}^{\theta_+} C_x \frac{dC_y}{d\theta} d\theta.
\label{eq:geoarea}
\end{equation} 

%\section{References}

\end{document}